\documentclass[12pt]{amsart}

\usepackage{hyperref}

\usepackage{DLdef1}
\let\ssec\subsection

 \def\mmat #1,#2,#3,#4,{\text{\small\arraycolsep=3pt $
\begin{pmatrix}#1&#2\\#3&#4\end{pmatrix}$}}

\begin{document}

\title{Two  problems in the theory of  differential equations}

\markboth{Dimitry Leites}{Two  problems in the theory of DE}

\author{D. Leites}

\address{New York University Abu Dhabi, Division of Science and Mathematics, P.O. Box 129188, United Arab
Emirates, dl146@nyu.edu; Department of mathematics, Stockholm University, Roslagsv.
101, Kr\"aft\-riket hus 6, SE-106 91 Stockholm, Sweden\\
mleites@math.su.se}

\begin{abstract} 1) The differential equation considered in terms of exterior
differential forms, as \'E.Cartan did, singles out a differential ideal in the supercommutative superalgebra
of differential forms, hence an affine
supervariety. 
In view of this observation, it is evident
that every differential equation has a supersymmetry (perhaps trivial).  Superymmetries of which (systems of) classical differential equations are missed yet?

2) Why criteria of formal integrability of differential equations are
never used in practice? 
\end{abstract}



\maketitle

\markboth{\itshape Dimitry Leites}{{\itshape Two problems in the theory of differential equations}}

\begin{quote}\rightline{To the memory of Ludvig Dmitrievich Faddeev}\end{quote}

\thispagestyle{empty}

\section{Introduction}\label{Intro}

While the Large Hadron Collider is busy testing (among other things) the existence of supersymmetry in the high energy physics (more correctly, supersymmetry's applicability to the models), the existence of supersymmetry in the solid body theory (more correctly, its usefulness for describing the models) is proved beyond any doubt, see the excitatory book \cite{Ef}, and later works by K.~Efetov. 

Nonholonomic nature of our space-time, suggested in models due to Kaluza--Klein (and seldom remembered Mandel)  was usually considered seperately from its super nature. In this note I give examples where the supersymmetry and non-holonomicity manifest themselves in various, sometimes unexpected, instances. In the models of super space-time, and super strings, as well as in the problems spoken about in this note, these notions appear together and interact. 

L.D.~Faddeev would have liked these topics, each of them being related to what used to be of interest to him. Many examples will be only mentioned in passing. 

\ssec{History: Faddeev--Popov ghosts as analogs of momenta in \lq\lq odd" mechanics} Being elected Member of the USSR Academy of Science, 
L.D.Faddeev got the right to submit papers to Soviet Mathematics 
Doklady, a rather prestigious journal at that time. The first paper 
he presented having become Academician was my note \cite{L2} in which I gave not only an 
interpretation of the Schouten bracket (lately often called \textbf{antibracket}, and sometimes \textbf{Buttin bracket}, in honor of the first person who proved that the antibracket satisfies the super Jacobi identity) as an analog of the 
Poisson bracket (this analogy was manifest from the moment the Schouten bracket was  discovered, but without important examples of application, like the equation  $\dot f=\{f, h\}_{a.b.}$ and its interpretation). I also interpreted the quotient of the Buttin superalgebra $\fb(n)$, on which the Schouten bracket is the multiplication, by the center 
as preserving an odd (in both senses) analog of the symplectic form. This quotient, $\fle(n)$, preserves an an odd analog of the symplectic form, and is an \lq\lq odd" superization of the Lie algebra $\fh(2n)$ of Hamiltonian vector fields,  whereas the Lie superalgebra $\fh(2n|m)$ --- the quotient by the center of the Poisson Lie superalgebra $\fpo(2n|m)$, ---preserving an even (ortho)symplectic (nondegenerate and closed) even 2-form is a \lq\lq direct" superization.

This latter interpretation was new.

The Lie superalgebras $\fle(n)$, the corresponding \lq\lq odd" analog $\fm(n)$ of the Lie algebra of contact vector fields, preserving a distribution with an  \lq\lq odd time"\ $\tau$ singled out by the Pfaff equation with the form
\be\label{FP}
d\tau-\sum \pi_idq_i,
\ee
together with the divergence--free analogs of these superalgebras were the first of several series of counterexamples 
to a \lq\lq Theorem" and a Conjecture in \cite[Th.10, p.92; Conj. 1, p.93]{K}, claiming the classification of simple (and even primitive) Lie superalgebras of vector fields. Grozman found a deformation of the Schouten bracket and more counterexamples, see \cite{GrI}, but his result was appreciated only much later, see \cite{LSh5}.

It then took 2 decades to formulate the classification of \textit{simple vectorial} Lie superalgebras without mistakes, and without claiming having solved a wild problem of classification of \textit{primitive} Lie superalgebras as in \cite{K, K7}, see \cite{Sh14}; it took 10 more years to completely 
prove it, see \cite{LSh5}, and \cite{K7} with numerous rectifications, see, e.g.,  \cite{CaKa2, CaKa4, CK1a, CCK}. Observe that classification of deforms (the results of deformations) of simple vectorial Lie superalgebras listed in \cite{LSh5}, especially with odd parameters, was never published to this day, bar occasional examples, see  \cite{LSh3}, whereas the ones rediscovered in \cite{CK1}, were discarded by the authors. People (even mathematicians) still look askance at the odd parameters of deformations, although these parameters constitute the very tsimes of supersymmetry, whereas the odd central extensions are welcome, despite the likeness of cohomology that describes each of them.

The note \cite{L2} also corrected the formulation of the super version of 
Darboux's theorem (on the normal shape of the non-degenerate closed differential 2-form) in \cite{Kos}, where odd forms were not even considered.

Most importantly, \cite{L2} proclaimed the existence of an \lq\lq odd" analog of mechanics. Speaking at that time about the two analogs of symplectic forms --- and ensuing mechanics --- I used to joke that I could not believe that
God blessed only the Poisson one and not both of them.

In 1976--77, M.~Marinov told me, leaving theology for future (which did not hesitate to come),  that unless I demonstrate how to quantize this \lq\lq odd" mechanics, no physicist will listen to me, especially since the \lq\lq Planck's constant" for the \lq\lq odd" mechanics, if exists, should be odd, which is so odd he did not believe it could be so. 

Marinov was too pessimistic. This  \lq\lq odd" mechanics, soon rediscovered by I.Batalin and G.Vilkovisky, drew a huge interest of physicists, thanks to applications demonstrated already in their first paper  \cite{BV}. This happened despite the fact that the \lq\lq odd" mechanics were quantized much later (and to an extent; I think this is still an open problem since it is unclear how to interprete multiple analogs of Fock spaces and work with them, see  \cite{LSh3}). 

In any case, \lq\lq Faddeev--Popov's ghosts" got an interpretation (e.g., as \lq\lq coordinates" $\pi_i$ in eq.~\eqref{FP}). A bit later, mathematicians got interested in derived of the Schouten construction, called BV-algebras; among numerous works, let me point at a recent one: \cite{DSV}. 

\ssec{Towards future} In this note I want to point at two open problems, or
rather topics for research, in the area of differential equations.
More or less implicitly, these problems were being observed for
decades, but their formulations was published only sketchily and only in Russian. Clearly, there are more than two problems of importance in the
theory of differential equations; my choice of \textbf{the} two
problems discussed below reflects only my taste and understanding, not
claiming for more.

The problems to be discussed here, although important,  were beyond reach until recently for the lack of adequate tools and even terms, so these
problems did not draw any attention, to say nothing about these
attention they deserve in my opinion. Now the time is more propitious for solving these problems: we have at our
disposal not only the vocabulary but also a technique which,
speaking of the second topic, is implemented, to an extent, in a
\textit{Mathematica}-based package  \cite{Gr}.

To make presentation reasonably short (as a
speach at the wake), all the excruciating details are replaced by references to illuminating text-books and monographs \cite{BCG,
Del, LSoS, Ber}, and papers \cite{L3, Shch,Sha}.

For a criticism of epygony and wrong texts that should not have been published, see one of Appendices to the paper \cite{Mo}, devoted to infinite-dimensional supermanifolds, cf. Wheeler's conjecture in epigraph to the next section.

\section{Supersymmetries of differential equations}\label{SUSY}

\begin{quotation}\lq\lq The stage on which the space of the Universe moves is certainly not space itself. Nobody can be a stage for himself; he has to have a larger arena in which to move. The arena in which space does its changing is not even the space-time of Einstein, for space-time is the history of space changing with time. The arena must be a larger object: \textit{Superspace} \dots It is not endowed with three or four dimensions --- it is endowed with an infinite number of dimensions.\rq\rq (J.A. Wheeler: \textit{Superspace}, Harper's Magazine, July 1974, p. 9, see \cite{Giu}.)\end{quotation}

\ssec{Sophus Lie, \'Elie Cartan and supersymmetry}\label{Lie} As is well-known, 
Sophus Lie introduced
the groups, now bearing his name, in order to carry out for
differential equations the analog of what Galois did for algebraic
equations. Lie's work was the beginning of the systematic study of
symmetries of differential equations and their solutions.

Several monographs and text-books,  and countless research papers
are devoted to the study symmetries of
differential equations.

Nobody, however, observed (at least, this observation was never made
public, except briefly in \cite{LSoS}) that every differential
equation  possesses a \textbf{super}symmetry  (SUSY
for short, as physicists say), and hence nobody tried to describe
this SUSY. (I am speaking about DE on any manifold, not on a supermanifold where a SUSY, but not the one I am now writing about, is present by definition.)

Since ca 1974,
i.e., shortly after I've understood how to define what in modern terms
is called \textit{affine superscheme} or \textit{supervariety}, see
\cite{L1}, I was planning to rectify this oversight. 

The discovery of supersymmetries by J.~Wess and B.~Zumino
explained that the Maxwell equation and the Dirac equation
should not be considered as separate equations, the Lie algebra of
symmetries of each of them being the Poincar\'e algebra (see, e.g., \cite{FN}). They should be considered as
a system of equations tied by a supersymmetry whose Lie superalgebra
is currently called $N=1$ Poincar\'e superalgebra, see \cite{DBS, BGLS} where the extensions for $N>1$ are described. This discovery
made my observation, described a bit further in this section, obvious, I
thought. Obviously, it was not as obvious as I thought it was; for
almost 40 years it was never rediscovered.

Later, I intended to delegate the problem of finding the
supersymmetry of classical equations of mathematical physics to
somebody among my students, but failed to find a sufficiently
interested one. Perhaps, my own enthusiasm was insufficiently contagious, or the
problem looked too vague, or time was not ripe yet. 

The current
occasion seems to me appropriate to make researchers interested in
this problem, and somewhat related problems I will describe in \S~\ref{Nonhol}.

\'E.~Cartan performed a step most important from super point of
view: he reformulated the notion ``differential equation" in terms
of exterior differential forms. For an exposition of this approach,
see the books \cite{BCG} and \cite{C}.

For simplicity, consider an ODE: the passage to PDE, and to systems
of equations is evident. Recall that to a given differential
equation of order $k$ for an unknown function $u$ depending on $x$,
i.e., to the expression of the form 
\begin{equation}\label{Ô}
F\left(x, u, \frac{du}{dx}, \dots,
\frac{d^ku}{dx^k}\right)=0,\end{equation} one can assign a
differential ideal (i.e., an ideal closed with respect to the
exterior differential $d$) in the superalgebra of functions in even (commuting) indeterminates $x, u,
p$, and odd (anti-commuting) indeterminates $dx, du, dp$, where $p=(p_1, \dots, p_k)$ and $dp=(dp_1, \dots,
dp_k)$, generated by the function $F(x, u, p)$ and exterior (odd,
anti-commuting) forms
\begin{eqnarray*}\label{123}
\renewcommand{\arraystretch}{1.4}
\omega_0&=du-p_1dx,\\
\omega_1&=dp_1-p_2dx,\\
\ldots&\ldots\ldots\ldots\ldots\ldots\\
\omega_{k-1}&=dp_{k-1}-p_kdx.
\end{eqnarray*}

The only thing that \'E.~Cartan did not do was to say \textbf{what}
does this differential ideal single out and where. To be able to say
this, he needed the definition of supervarieties (given in 1972, published in \cite{L1}), and basics of
supersymmetries, see lectures by J.~Bernstein (notes taken by
P.~Deligne and J.~Morgan) in \cite[pp.41--96]{Del} and a more detailed
text-book \cite{LSoS}. Equipped with this knowledge we can say: the
ideal in question singles out a certain sub\textbf{super}scheme or
sub\textbf{super}variety in the affine superspace whose algebra of
functions is the supercommutative
superalgebra  with even (topological) generators (\lq\lq coordinates") $x,
u, p$ and odd ones $dx, du, dp$.

There are several text-books teaching us the art of seeking
symmetries of differential equations, e.g., see the book \cite{FN} and 
references therein. Superizations of certain facts
and notions of algebraic and differential geometry are rather
intricate, see \cite{LSoS} and \cite{Sha}; in this particular case,
however, superization of the technique describing symmetries of differential equations seems to be
straightforward. Therefore, we are fully equipped for solving the
following problem.

\sssbegin{Problem}\label{P1} Describe supersymmetry groups (infinitesimally:
Lie superalgebras) of differential equations of mathematical
physics. Which of them do not reduce to groups (Lie algebras)? 
\end{Problem}

\begin{Remarks} 1) Physicists
(pioneers of supersymmetry) discovered supersymmetries that
intermix vector and spinor fields together with space-time
coordinates. In the light of the above interpretation of any differential
equation as something that singles out a subsupervariety, this discovery does not look as
astounding as it is still represented in the literature. However, these physicists were the first to observe SUSY and this is what they are being praised for. Remarkably, some of these pioneers do not understand the importance of their own discoveries even now, see their prefaces in \cite{DBS}. 

2) It was Wess and Zumino who used the catchy term \textit{supersymmetry}. Soon after, Salam and Strathdee  termed the object, on which supergroup of supersymmetry acts --- \textit{superspace} --- as is now customary to call the pair $\cM=(M, \cO_{\cM})$ consisting of the space (manifold) $M$ ringed by the sheaf of supercommutative superalgebras $ \cO_{\cM}$. Two years earlier, in the purely algebraic setting, where the only functions considered are polynomial and rational, I used for what is now called \textit{superscheme} or \textit{algebraic supervariety} the dull and boring term  \textit{spectrum of graded-commutative ring}, see \cite{L1}\footnote{\lq\lq From somewhere in the 1950s on, John Wheeler  repeatedly urged people who were interested in the quantum-gravity program  to understand the structure of a mathematical object that he called \textit{Superspace} \cite{W}. The intended meaning of \textit{Superspace} was that of a set, denoted by $S(\Sigma)$, whose points faithfully correspond to all possible Riemannian geometries on a given three-manifold $\Sigma$\rq\rq, see \cite{Giu}. Thus, Weeler (known not only as Feynman's teacher, but also for having coined the terms \lq\lq black hole" and \lq\lq wormhole") used the term \lq\lq superspace" in a sense absolutely different from what is now customary to use by everybody, bar geometro-dynamists. 

It is a unfortunate that the same term has two completely different meanings because both meanings of the term can meet in one sentence, e.g., when the points of Wheeler's \lq\lq superspace" are generalized to consist of super-Riemann manifolds.}.\end{Remarks}

\sssec{Gauge fields from super point of view} Let $\cF$ be the sheaf (this is not yet popular term among the physicists; locally, it is the algebra) of functions on $\cM$, and $V=\Gamma(\cM, \cE)$ is the $\cF$-module of sections of the bundle $\cE$. On the \textbf{level of points}, a \textit{connection} on a vector bundle $\cE$ over a (super)manifold $\cM$ is an odd map $\nabla\colon V\tto \Omega^1(\cM)\otimes_\cF V$, by the formula 
\[
\nabla(fv)= df \otimes v+ (-1)^{p(f)}f\nabla(v) \text{~~ for any $f\in\cF$ and $v\in V$}.
\]
Having fixed a \textit{flat} connection $\nabla_0:=d$, meaning here, by the usual abuse of notation, $d\otimes 1_{\dim V}$ rather than $d$, we can represent any connection in the shape 
\be\label{formNab}
\nabla=d+\alpha, \text{~~ where $\alpha\in (\fgl(\rk V)\otimes\Omega^1(\cM))_\od$.}
\ee
The form $\alpha$ is called \textit{the form of $\nabla$} or a \textit{gauge field} with gauge group $\GL(\rk V)$ or any smaller sub(super)group $G\subset \GL(\rk V)$, if $\alpha\in (\fg\otimes\Omega^1(\cM))_\od$ for the Lie (super)algebra $\fg$ of $G$. 

The form $F_\nabla=\nabla^2\in (\fg\otimes\Omega^2(\cM))_\ev$ is called the \textit{curvature form} of the connection $\nabla$ or the \textit{stress tensor} of the field $\alpha$.

\paragraph{Super specifics, unexplored to this day, cf. \cite{MaG, Del}} 1) Observe that the definitions in the above paragraph describe only \textbf{points} of the linear supermanifold of connections with the gauge superalgebra $\fg$ on $\cM$ and the \textbf{points} of the linear supermanifold of curvature forms of these connections. To describe the odd parameters of these supermanifolds  we use the functors of points (\lq\lq Grassmann envelopes" in Berezin's words, see \cite{Ber}) that to every supercommutative superalgebra $C$ assign the sets of $C$-points of connections $(\fg\otimes\Omega^1(\cM)\otimes C)_\od$ and curvature forms $(\fg\otimes\Omega^2(\cM)\otimes C)_\ev$.

2) The map $\nabla$ can be extended to the higher differential forms (sections of higher exterior powers of the cotangent bundle) and this is well-known.

On supermanifolds, additionally, $\nabla$ can be extended to the integrable forms (volume-valued sections of higher exterior powers of the tangent bundle), where the  $\cF$-module of integrable $i$-forms is defined to be $\Sigma_{-i}(\cM):=(\Omega^i(\cM))^*\otimes_\cF \Vol(\cM)$, where $\Vol(\cM)$ is the $\cF$-module of volume forms (aka Berezinian): 
\be\label{*}
\nabla\colon \Omega^i(\cM)\otimes_\cF V\tto\Omega^{i+1}(\cM)\otimes_\cF V; \ \
\nabla\colon V\otimes_\cF\Sigma_j(\cM)\tto V\otimes_\cF\Sigma_{j+1}(\cM).
\ee
Let 
$
T\colon\Omega^i(\cM)\otimes_\cF V\simeq V\otimes_\cF  \Omega^i(\cM)$
be a twisting isomorphism 
\be\label{**}
T(\omega \otimes v)=(-1)^{p(\omega)p(v)}v\otimes \omega\text{~~ for any $\omega\in \Omega^i(\cM))$ and $v\in V$}
\ee
and $\alpha$ be defined in~\eqref{formNab}. We define $\nabla$ in eq.~\eqref{*} by means of the following formulas  
\be\label{dop}
\begin{array}{l}
\nabla(\omega\otimes v)=d\omega\otimes v+(-1)^{p(\omega)}\omega\otimes \alpha(v) \text{~~ for any $\omega\in \Omega^i(\cM))$ and $v\in V$}; \\
\nabla(v\otimes \sigma)=T(\nabla(v))(\sigma)+(-1)^{p(v)}v\otimes d\sigma \text{~~ for any $\sigma\in \Sigma_j(\cM))$ and $v\in V$}. \\
\end{array}
\ee

Observe that   
\begin{equation}\label{deRh}
\begin{minipage}[l]{14cm}
\textbf{The de Rham cohomology of any supermanifold is isomorphic to the de Rham cohomology  of the underlying manifold}, \text{see \cite{MaG}.}\end{minipage}
\ee

Let \textit{pseudo}forms be a common name for various  types of smooth functions in coordinates $x_i$ and their differentials $\widehat x_i:=dx_i$. There are several types of pseudoforms: of rapid descent at infinity, homogeneous (such that $f(x, t \hat x)=t^a f(x,  \hat x)$ for any $t\in \Ree$ and some $a\in\Ree$), etc. 
\begin{equation}\label{PseudodeRh}
\begin{minipage}[l]{14cm}
\textbf{The pseudo(co)\-homology on the space of pseudoforms is NOT isomorphic, generally, to the de Rham cohomology  of the underlying manifold}, \text{see \cite{Zo}.}
\end{minipage}
\end{equation} 
The fact \eqref{PseudodeRh} is most interesting  since $\nabla$ can be extended by the operators given by eqs. \eqref{*} --- \eqref{dop} from the space of forms, which are polynomials in the (even) differentials of the odd coordinates,  to the  space of \textit{pseudoforms}.

Given two $\cF$-modules with connections $(V_i, \nabla_i)$, where $i=1,2$, define their tensor product $(V_1\otimes_\cF V_2, \nabla_{1\otimes 2})$ and the module of homomorphisms $(\Hom(V_1,  V_2), \nabla_{1^*\otimes 2})$ by setting for any $v_i\in V_i$ and $F\in \Hom_\cF(V_1, V_2)$:
\[
\begin{array}{l}
\nabla_{1\otimes 2}(v_1\otimes v_2)=\nabla_{1}(v_1)\otimes v_2+(-1)^{p(v_1)}(T\otimes 1)v_1\otimes \nabla_{2}(v_2) \text{~~ for any $\omega\in \Omega^i(\cM))$ and $v\in V$}; \\
\nabla_{1^*\otimes 2}(F)(v_1)=\nabla_{2}(F(v_1)))-(-1)^{p(F)}(1\otimes F)(\nabla_{1}(v_1)). \\
\end{array}
\]
If $V_1=V_2=V$, we set 
\be\label{end}
(\End(V), \nabla^{\End}):=(\Hom(V_1,  V_2), \nabla_{1^*\otimes 2}).
\ee

3) One more aspect of super specifics is that the physically meaningful supermanifolds, e.g., $N$-extended Minkowski superspace for any $N$, and each of the superstrings considered \textit{usually}, is endowed with a \textbf{non-integrable} distribution and is neither real, nor complex supermanifold but a \textit{real supermanifold with a complex structure on subspaces that define the distribution}, see \cite{BGLS}. 

The connection on such \textit{real-complex supermanifold} is what is called \textit{circumcised} (or \textit{reduced}) connection; the form of such a connection is defined by its values on the subspaces that define the distribution. 

4) Quillen was the first to apply \lq\lq superconnections" to prove the index theorem, see \cite{Q1} and a later exposition in the book \cite{BGV}. Quillen's proof was based on the isomorphism spoken about in \eqref{deRh}. For details of an open problem \lq\lq how to superize the index theorem", see \cite{Li}.

\sssec{Well-known example: SUSY of Maxwell and Dirac equations} Let $M$ be a manifold locally diffeomorphic to the Minkowski space with coordinates we arrange  in a $2\times 2$ matrix $(x_{ab})=\begin{pmatrix}x_0-x_3&x_2-ix_1\\ x_2+ix_1&x_0+x_3\end{pmatrix}$, where $x_0, \dots, x_3$ are the usual local coordinates in the Minkowski space with the metric $\det(dx_{ab})$, where $a, b=0,1$ and the differentials $dx_{ab}$ are considered even (commuting). 

Let $A=\{A_\mu(x)\mid \mu=0,\dots, 3\}$ be the vector-potential of an electro-magnetic field. For mathematicians, the coordinates of this $A$ are components of  the form of the connection (a.k.a. gauge field) $\nabla:=d+\sum A_\mu dx_\mu$, where the differentials are considered odd (anti-commuting). Let $F_\nabla=\sum F^{\mu\nu} dx_\mu\wedge dx_\nu$, and $\ast F_\nabla$ its Hodge\footnote{Recall the definition. For any non-degenerate symmetric bilinear form of signature $(s, n-s)$, i.e., with $s$ minuses in the diagonal normal shape of (the Gram matrix of) $g=\diag(-1, \dots, -1, 1, \dots, 1)$,  on the $n$-dimensional (co)tangent space to the manifold $M$, we set $g(1,1)=1$ and extend $g$ from $\Omega^1_c$, where $g(dx_i, dx_j)=\pm \delta_{ij}$ with the first $s$ diagonal values negative and the rest positive,  to the space $\Omega^p_c$ of differential $p$-forms (with constant coefficients, as indicated by the subsript), by setting 
\[
g(\omega_1, \omega_2)=\det (g(dx_{i_k}, dx_{j_l})_{k,l=1}^p),\text{~~ where $\omega_1=dx_{i_1}\wedge\dots\wedge dx_{i_p}$, $\omega_2=dx_{j_1}\wedge\dots\wedge dx_{j_p}$}.
\] 
For any form $\omega\in \Omega^p_c$, there is a unique form $\ast\omega\in \Omega^{n-p}_c$, called its \textit{Hodge dual}, such that 
\[
\omega\wedge\ast\omega=(-1)^sdx_1\wedge \cdots\wedge dx_n. 
\]
This definition is naturally extended to the whole space $\Omega^{\bcdot}=\oplus \Omega^{p}$ with any coefficients.} dual. The \textit{Maxwell equation} is the simplest  \textit{Yang-Mills equation}, the one with the smallest (1-dimensional) gauge group for the case where $\rk V=1$.

Namely, let $\psi:=\{\psi^a\mid a=0,1\}$ and $\overline \psi:=\{\overline \psi^{b}\mid b=0,1\}$ be the spinor wave-function and its Dirac-adjoint, i.e., $\psi\in\Cee^2$ on which $\fo(3,1)\simeq\fsl_\Cee(2)$ acts via the \textit{spinor representation}\footnote{The notion is well-known, of course, but its most lucid description --- in terms of Lie superalgebras, see \cite{How, Ber} --- is not  yet a common knowledge, regrettably.}, and  $\overline \psi\in(\Cee^2)^*\simeq\Cee^2$ is an element of the \textit{hermitian dual} (dual and complex conjugate) spinor representation. The \textit{Maxwell equation} with the source given by an electron-positron field  is
\be\label{Max}
\nabla^{\End}(\ast F_\nabla)=\psi\otimes \overline \psi\text{~~(for the definition of $\nabla^{\End}$, see eq. \eqref{end})}.
\ee

Let $\hbar=1$ and $c=1$ whereas $e$ and $m$ are the charge and mass of the electron, respectively. The \textit{Dirac equation} for the electron in the electro-magnetic field can be written  as
\be\label{Dir}
\sum_\mu \gamma^{\mu}(i\partial_\mu+eA_\mu)\psi=m\psi,\text{~~where $\gamma^{\mu}$ are the Dirac matrices},
\ee
and similarly for the Dirac-adjoint spinor $\overline \psi$. The symmetry group of each of the equations \eqref{Max} and  \eqref{Dir} is the same --- the Poincar\'e group, i.e. $\text{O}(3,1)\ltimes\Ree^4$, see, e.g., \cite{FN}.

Now, consider the tangent space to the Minkowski superspace $\cM_1$ с координатами $(x_{ab}, \psi^a, \overline \psi^{b})$, где $a, b=0,1$, realized as a Lie subsuperalgebra of one of the real forms of the complex Lie superalgebra $\fsl(4|1)$ by supermatrices in the nonstandard format 
 $\fsl(2|1|2)$, and vector fields, as follows, see \cite[eq.~(I), p.3, and p.231]{WB}:
\be\label{tMink}
\begin{pmatrix}
0&0&0\\
Q&0&0\\
T&\overline Q&0\\
\end{pmatrix}, \text{~~where $Q_a=\partial_{\psi^a}+\overline\psi^{b}\partial_{x_{ab}}$ and $\overline Q_{b}=\partial_{\overline\psi^{b}}+\psi^{a}\partial_{x_{ab}}$, $T_{ab}=\partial_{x_{ab}}$}.
\ee
Let $\sigma_{\mu\nu}:=\frac{i}{2}[\gamma^\mu,\gamma^\nu]$ and let $\eps^a$ and $\overline\eps^{b}$ are the odd parameters forming 2-component spinors, like $\psi$ and $\overline \psi$, respectively. The direct verification shows that the transformations generated by odd generators $Q_a$ and $\overline Q_{b}$ act as follows (only the linear part of the increment is shown): 
\[
(\psi, \ \overline\psi, \ A)\longmapsto \left(\sigma_{\mu\nu}F^{\mu\nu}\eps,\  -\overline\eps\sigma_{\mu\nu}F^{\mu\nu}, \
i(\overline\eps\gamma^\mu\psi-\overline\psi\gamma^\mu\eps)\right).
\] 
 Hence,
\begin{equation}\label{susy}
\begin{minipage}[c]{14cm} 
\textbf{the symmetry of the system, consisting of \eqref{Max}, and \eqref{Dir} with its Dirac-adjoint, is the Lie subsuperalgebra of $\fsl(2|1|2)$ whose elements are supermatrices of the form}
\[
\begin{pmatrix}A&0&0\\
Q&0&0\\
T&\overline Q&-\overline A^T\\
\end{pmatrix},  \text{~~where $A\in\fsl_\Cee(2)$}.
\]
\end{minipage}
\end{equation}

For a description of $N$-extended Minkowski supermanifolds $\cM_N$, and related structures, like the \textit{curvature tensor taking into account the non-integrable distribution} on $\cM_N$, see \cite{BGLS}.

\sssec{Witten's discoveries: two more types of SUSY related to differential equations} These SUSYs are close
to those considered above but differ from them.

1) Witten's interpretation of integrability criteria of Yang--Mills equations, see \cite{W1} and an extensive comment in the book \cite{MaG}. 

2) SUSY of  the ``conventional"
Schr\"odinger equation with matrix potential, see \cite{W2}, see a lucid review \cite{GK} which inspired numerous works with keyword \lq\lq shape invariance". This SUSY explains degeneration of the non-zero spectrum of the equation and is a realization (representaion) of $\fq(1)$, a~\textit{queer} analog of $\fgl(1)$. {For an explanation why this Lie superalgebra is an analog of $\fgl(n)$, see \cite{LSoS}; for a description of irreducible representations of $\fq(n)$, see \cite{Br, ChK}.)

\sssec{Kirillov's observation: a SUSY-related interpretation of a known property of solutions of the Sturm--Liouville equation} A.A.Kirillov noticed, see \cite{Kir},  that the known fact \textit{the product of any two solutions of the Hill (Sturm--Liouville) equation
\[
2y''+p(x)y=0
\]
is a solution of the equation} 
\[
y'''+2p(x)y'+p'(x)y=0
\]
follows from the description of the stationary Lie superalgebra of the point in the co-adjoint representation of the Neveu--Schwarz Lie superalgebra, one of the 12 non-trivial central extensions of simple stringy (aka superconformal) Lie superalgebras, see \cite{GLS}. A similar investigation of the stationary Lie superalgebras of the points in the co-adjoint representations in 9 of the remaining 11 cases, see \cite{OOCh, Lkur}.

\section{Formal integrability of differential systems and non-holonomic
structures}\label{Nonhol} 

In \cite{BCG} one can find, among other things, criteria of
formal integrability of differential equations. 

There are  several known ways to solve a given DE or systems of DE; the method offered in \cite{BCG} differs from the other methods  I heard  about in that it is NEVER
used in practice (by engineers, physicists, chemists, biologists, geologists, etc.). It is still used sometimes in grant applications  by mathematicians, but this is hardly what taxpayers  usually  have in mind  speaking about ``applications" of the method.  

It seems to me that the trouble with the  problems discussed in the book \cite{BCG} was the same as with
the problem described in \S~\ref{SUSY}: time was not ripe yet, several vital notions were lacking.

\ssec{Criteria for formal integrability of differential equations}
These criteria are expressed in \cite{BCG} in terms of the Lie
algebra $\fg$ of local symmetries of the DE and the certain
\emph{Spencer cohomology} of $\fg$, where $\fg$ is supposed to be a
$\Zee$-graded Lie algebra $\fg=\mathop{\oplus}_{i\geq
-d}\fg_i$ of depth $d=1$. Among the manifolds with a $G$-structure, the simplest ones are \textit{flat}, see \cite{BGLS}. The cocycles representing nontrivial
classes of the same Spencer cohomology are the obstructions to
flatness of the $G$-structure, where $G$ is the Lie group whose
Lie algebra is the $0$th component $\fg_0$ of $\fg$. 

Examples of
such obstructions  well-known before they were expressed in terms of Spencer cohomology (note that examples (a) and (b) are meaningful over any field of characteristic $\neq 2$): 

(a) the Riemann tensors for metrics, here $\fg=\fo(n)$; 

(b) $d\omega$ for
the nondegenerate exterior 2-form $\omega$, here $\fg=\fsp(2m)$; 

(c) the Nijenhuis tensor for
a given almost complex structure, here $\fg=\fgl_\Cee(m)\subset\fgl_\Ree(2m)$.)

\textbf{Question No.1}: why nobody among all those who solve
DEs every day
--- ever use this approach and use instead numerical approximations
and scores of other methods?

Various methods work under different conditions; e.g., it is more reasonable to approximate $\sin x$ by the Fourier series, not Taylor one, but even the latter works in a neighborhood. It is very difficult to imagine a method that NEVER works. So, when and for which DE should one use Spencer cohomology as means for getting a solution?

Although
essentially all method for solving a given (system of) DE (should) lead to the same answer, computing (co)homology is at the moment not
efficiently organized. Even the most efficient code known to me
(\emph{SuperLie}, see \cite{Gr}) is applicable to a small portion of the Lie algebras whose (co)homology we'd like to compute, cf.  \cite{MF}; for a comparison of \emph{SuperLie} efficiency with  an \textit{ad hoc} code written by a professional, see \cite{Kor}).

It seems strange to me that although to \textbf{define} Lie algebra
cohomology $H^i(\fg; M)$ is possible very succinctly, by means of
structure constants of the actions of $\fg$ in the adjoint module and the module of
coefficients $M$, \textbf{computing} cohomology is --- at the moment
--- a task formulated everywhere so clumsily (from the point of view of the programmer and computer) that the time needed
grows exponentially with $\dim \fg$, $\dim M$ and $i$. One has to use
the same data again and again. It is hopeless to  directly compute $H^i(\fg; M)$ starting with $\dim \fg>15$ and $i>4$ using any of the  widely known codes/platforms. 

\textbf{Is the complexity of computing (co)homology of the Lie algebra $\fg$  in-built indeed, and the volume of computations grows exponentially with $\dim \fg$ and $i$}? (We'd like to compute, e.g., $H^i(\fg)$ for $\dim \fg<300$ and $i<4$.)

\textbf{Question No.2}: According to a
well-known theorem, the symmetries of a given DE are induced by
either point or contact transformations. Why the integrability criteria of DE given in
\cite{BCG} and later works, only deal with  DEs of the first type.

The answer to \textit{this} question is clear to me: there are several snags.

First, the symmetry algebras, are, strictly speaking, filtered, not graded.

Second, the integrability criteria given
in \cite{BCG}  are expressed in terms of \emph{Spencer cohomology}, see \cite{Po}.

Until recently, this cohomology was only defined for the $\Zee$-graded Lie algebras
$\fg=\mathop{\oplus}_{i\geq -d}\fg_i$ of depth $d=1$, whereas
the analogs of Spencer cohomology for the DE whose symmetries are induced from contact transformations should be computed for the graded
Lie algebra of depth $>1$ associated with filtered Lie algebra of symmetries of the given DE.

The definition of analogs of Spencer cohomology for the graded Lie (super)algebras  of any depth associated with a filtered one, see in 
\cite{GrL}, where there are considered  various versions of supergravity equations (SUGRA) for any $N\leq 8$, cf. with the best for today book about SUGRA, \cite{GIOS}, where SUGRA are deduced for $N\leq 3$. Regrettably, \cite{GrL} is written in the jargon alien to physicists and differential geometers: without Christoffel symbols but in terms of Lie superalgebra cohomology instead; that is why nobody, it seems, read the paper.

The difference between cohomology of the filtered and associated with it graded Lie superalgebra is controlled by what, in the particular case of $N=1$ SUGRA equations, physicists call \textit{Wess-Zumino constraints}.

To generalize the criteria for formal
integrability to embrace the DEs whose symmetries are induced by contact transformations, new notions and results are needed. Now we have them. 

First: we need the generalization of Cartan prolongs for Lie (super)algebras of depth $>1$, due to Shchepochkina \cite{Shch}. Instead, people often refer to N.Tanaka. It is instructive to compare \cite{Shch} with the earlier, but less useful, works by other authors, e.g., by N.~Tanaka
(not with the first one, where the square of the differential, by means of which one was supposed to compute cohomology, did not vanish, but with reasonable later ones), and other meaningful ones, e.g., \cite{Ze, AD}, in which \cite{Shch} is not even mentioned. The paper \cite{Shch} embraces the super case, and
characteristic $p>0$ case, and \emph{partial prolongs} --- all these for the first time, and gives the most convenient algorithm for realization of Lie (super)algebras by creation and annihilation operators (i.e., by vector fields, speaking prose).

Second: we need the definition of an analog of Spencer cohomology for Lie (super)algebras of depth  $>1$, see \cite{L3}.

\sssec{Hertz, \'E.Cartan, Carath\'eodori, V.Sergeev, and non-holonomic structures}\label{sss211} A nonintegrable \textit{distribution} (meaning a subbundle of the
tangent bundle, not a generalized function) is said to be
\emph{non-holonomic}. The criterion for integrability of a given
distribution is given by the well-known \textbf{Frobenius theorem}: \textit{the distribution $\cD$ on a (super)manifold $M$ is integrable if sections of $\cD$ form a Lie sub(super)algebra in the Lie (super)algebra of vector fields on $M$}.

The term  \textit{non-holonomic} was coined by H.~Hertz who considered
mechanical systems with linear constraints on velocities, such as a
ball on a rough plane, or any vehicle (at the point of tangency with
asphalt the wheel's velocity vanishes).

Examples of manifestation of non-holonomic nature of certain mechanical
systems are ubiquitous: from the cat's ability to land on its feet
having fallen out of the window (``Cat's problem"), to guided
missiles, to skates, to basket balls refusing to fall into the
basket; Internet returns tens of thousands of examples.

More generally, a \emph{non-holonomic manifold} (no dynamics) is the
one endowed with a nonintegrable distribution. A natural way to
define a distribution is to single it out as a set of zeros of a
system of \emph{Pfaff equations} whose left hand sides are
differential 1-forms. The simplest and most known non-holonomic
distribution is singled out by the Pfaff equation for vector
fields $X$, given by the contact form $\alpha$:
\[
\alpha (X)=0, \ \text{ where $\alpha=dt-\sum p_idq_i$}.
\]

Interesting examples of non-holonomic structures are provided by
modern high energy physics, e.g., \textit{Minkowski} superspaces and
superstrings \textit{with a contact structure}, see \eqref{susy} and\cite{BGLS} (whereas arbitrary superstrings or supermanifolds are not necessarily endowed with a non-holonomic distribution).

Note that the first definitions of the exceptional simple Lie algebras \'E.Cartan and Killing gave in terms of non-holonomic distributions these algebras preserve, see  \cite{Shch}, not in terms of Cartan matrices and diagrams bearing the name of Dynkin (who was not yet born then). Lately people started to return to the Cartan-Killing way of looking at the Lie algebras due to its applications in the study of PDEs, see \cite{The}.

In 1920s, C.~Carath\'eodory reformulated thermodynamics as a
non-holonomic mechanical model provided we do not require that the
temperature or energy of the gas is a constant.

In 1980s, V.~Sergeev rewrote several first
pages of the text-book by Landau and Lifshitz on statistical physics
in terms of market economy, see \cite{S}, with words ``market
agent" (seller/buyer) instead of ``particle". This is one of the first, if not
\textbf{the} first, monographs on what nowadays is called ``econophysics",
more interesting, in my opinion, than other books on the topic I
read so far. The main mathematical observation of Sergeev's book (for me; the book contains other results, e.g., deducing Chatelier's principle) is
the same as Carath\'eodory's: the heart of the matter lies in
non-holonomic nature of the object of the study (see also \cite{Pa, PS});  nobody yet studied markets \textbf{in this way}.

\ssec{Nonholonomic structures and the corresponding analogs of the curvature tensor}\label{sss211} For a long time nobody could define an analog of the curvature
tensor (field, more correctly say) that would take into account presence of a
non-holonomic distribution. In his review of mathematics of
non-holonomic dynamics studied together with L.~Faddeev (with
possible ramifications to other types of nonintegrable constraints,
such as in linear programming), Vershik even conjectured that, despite
several examples where such tensor fields were defined and computed,
there is probably no general definition, see Vershik's Appendix in
\cite{Seng} --- the translation of \cite{S}  with appendices.

In \cite{L3}, the definition of the non-holonomic curvature tensor
was, however, given together with a reformulation of \lq \lq Spencer
cohomology"\footnote{To compute Spencer
cohomology is a grievous job to perform: there are no general
theorems on such cohomology, whereas for Lie algebra cohomology
there are several  general theorems (well, at least one), and a powerful trick called \lq\lq spectral sequence", see \cite{F, FF}.
The theorem I have in mind states that
the toral (i.e., maximal diagonalizing) subalgebra of the Lie (super)algebra $\fg$ acts by zero on any (co)homology of $\fg$ with any 
coefficients (at least, in finite-dimensional cases). This theorem greatly simplifies computations, but with an infinite number even of finite-dimensional cases the theorem fails, cf. \cite{MF}.

In my opinion, the notion of Spencer cohomology
is ``the opiate of the masses" (das Opium des Volkes), as K.~Marx
would say quoting (without due
reference)  Marquis de Sade. (Having learned this by chance from Wiki, I now better understand the roots of Marx's teaching than during the compulsory courses of history of the communist party of the Soviet Union.)} in equivalent, but much more convenient terms of Lie
algebra cohomology. 

As a result, we get a generalization of criteria
for formal integrability of DEs, given in \cite{BCG} for the DE whose symmetries are
induced by point transformations, to all DEs. Regrettably, it is unclear how to use these criteria. Observe that the \textit{nonlinear} constraints are also quite real.

\textbf{Nonlinear} constraints on velocities are represented, e.g., by the car with cruise
control switched on (so the velocity vector runs over a sphere of
radius, say, 55 mph). To \textit{define} integrability of such distributions,
we have to consider infinite dimensional manifolds of ``curved
Grassmannians" whose points are submanifolds of given dimension passing through a
given point (see \cite{MaG}, where finite-dimensional super models of 
curved Grassmannians of $0|k$-dimensional subsupervariety in the $0|n$-dimensional linear supervariety ($k\leq n$) are briefly discussed).

\subsection*{Acknowledgements.} It is my pleasure  to thank NYUAD for extremely stimulating working conditions. I am thankful to MPIMiS (Max-Planck-Institute for Mathematics in Sciences) at Leipzig, the town where S.~Lie worked most of his life, and where
I worked as a Sophus-Lie-Professor in 2004--07 when the ideas of this text were first written down, for a most creative
environment. I thank SIGMA's referees for comments that
helped me to improve the initial version of this text.


\end{document}